\documentclass[aps,prl,preprintnumbers,twocolumn,groupedaddress,nofootinbib]{revtex4}

\usepackage[dvips]{graphicx}
\usepackage{color}
\usepackage{amsmath,amssymb,slashed}
\usepackage{hyperref}

\usepackage{changepage}

\usepackage{tikz}
\usetikzlibrary{calc} 
\usetikzlibrary{patterns,snakes} 
\usetikzlibrary{decorations.pathreplacing} 
\usetikzlibrary{decorations.markings} 
\usetikzlibrary{decorations.pathmorphing} 
\usetikzlibrary{positioning}
\usetikzlibrary{arrows.meta}

\newif\ifdraft
\drafttrue
\newif\ifpreprint
\preprinttrue

\def\spa#1.#2{\left\langle#1\,#2\right\rangle}
\def\spb#1.#2{\left[#1\,#2\right]}

\def\beq{\begin{equation}}
\def\eeq{\end{equation}}

\newcommand{\eq}{\begin{equation}}
\newcommand{\eqe}{\end{equation}}
\newcommand{\eqa}{\begin{eqnarray}}
\newcommand{\eqae}{\end{eqnarray}}

\newcommand{\ep}{\epsilon}
\newcommand{\lamb}{\lambda}
\renewcommand{\ap}{\alpha'}
\newcommand{\te}{\textrm}
\newcommand{\bea}{\begin{eqnarray}}
\newcommand{\eea}{\end{eqnarray}}
\newcommand{\dd}{\mathrm{d}}

\newcommand{\KN}{\textrm{KN}}

\newbox\charbox
\newbox\slabox
\def\s#1{{     
        \setbox\charbox=\hbox{$#1$}
        \setbox\slabox=\hbox{$/$}
        \dimen\charbox=\ht\slabox
        \advance\dimen\charbox by -\dp\slabox
        \advance\dimen\charbox by -\ht\charbox
        \advance\dimen\charbox by \dp\charbox
        \divide\dimen\charbox by 2
        \raise-\dimen\charbox\hbox to \wd\charbox{\hss/\hss}
        \llap{$#1$}
}}

\begin{document}

\preprint{UUITP--17/21}
\preprint{NORDITA 2021-030}

\title{
Scattering Massive String Resonances through Field-Theory Methods
}

\author{Max Guillen$^a$, Henrik Johansson$^{a,b}$, Renann Lipinski Jusinskas$^c$ and
Oliver Schlotterer$^{a}$}
\affiliation{$^a$ Department of Physics and Astronomy, Uppsala University, \\ Box 516, 75120 Uppsala, Sweden}
\affiliation{$^b$ Nordita, Stockholm University and KTH Royal Institute of Technology,\\ Hannes Alfv\'{e}ns v\"{a}g 12, 10691 Stockholm, Sweden}
\affiliation{$^c$ Institute of Physics of the Czech Academy of Sciences \& CEICO \\ Na Slovance 2, 18221 Prague, Czech Republic}

\begin{abstract}
We present a new method, exact in $\alpha'$, to explicitly compute string tree-level amplitudes involving one massive state and any 
number of massless ones. This construction relies on the so-called twisted heterotic string, which admits only gauge multiplets, a gravitational 
multiplet, and a single massive supermultiplet in its spectrum. In this simplified model, we determine the moduli-space integrand of all amplitudes 
with one massive state using Berends--Giele currents of the gauge multiplet. These integrands are then straightforwardly mapped to gravitational 
amplitudes in the twisted heterotic string and to the corresponding massive amplitudes of the conventional type-I and type-II superstrings.
\end{abstract}

\maketitle

\section{Introduction}

\vspace{-0.3cm}
\noindent
The historical origin and the discovery of key features of string theory can be attributed to the study of its scattering amplitudes. Computations and structural properties of string amplitudes rely on exactly solvable correlation functions of vertex operators in a two-dimensional conformal field theory (CFT). For closed strings, the CFT approach leads to a factorization of the correlators into holomorphic and antiholomorphic building blocks, so-called {\it chiral correlators}. 
This property underlies the tree-level double-copy relation between perturbative 
gravity and gauge theories obtainable from string theory~\cite{Kawai:1985xq, Bern:2008qj, Bern:2019prr}, and inspired loop-level generalizations~\cite{chisplitt, Bern:2010ue}.

While the tree-level CFT prescription has long been textbook material \cite{Green:1987sp,Polchinski:1998rq}, recent discoveries of powerful double-copy structures {\it within the chiral correlators} have dramatically changed our perspective. Tree-level amplitudes of $n$ massless states of the open superstring \cite{Mafra:2011nv, Zfunctions} and the open bosonic string \cite{Huang:2016tag, Azevedo:2018dgo} can be factorized into {\it scalar} integrals over moduli spaces of punctured disk worldsheets and quantum field theory (QFT) building blocks carrying all the dependence on the external polarizations. In hindsight, this striking structure can be traced back to a decomposition of chiral correlators into a basis of integrals in the twisted cohomology defined by the moduli-space integration \cite{Mizera:2017cqs, Mizera:2017rqa}.  This {\it cohomology decomposition} is a general feature of string theory, its applicability to massless closed-string amplitudes was demonstrated in \cite{Schlotterer:2012ny, Stieberger:2013wea, Stieberger:2014hba, Azevedo:2018dgo}.

In this letter, we present the first all-multiplicity instance of double-copy structures
and cohomology decompositions of string amplitudes with \textit{massive} external 
states. More specifically, we describe a simple QFT setup that computes
the necessary building blocks for open and closed superstring amplitudes with $n{-}1$ massless and a single
massive level-1 state.
These are derived from Feynman diagrams of 10D super-Yang-Mills (SYM) theory deformed by a cubic operator involving two gauge multiplets and one spin-2 multiplet analogous to the first massive level of the open superstring.

Our QFT construction stems from the heterotic version of the {\it chiral} or {\it twisted string theories}~\cite{Hohm:2013jaa,Huang:2016bdd}. They differ from conventional strings by a relative sign flip of the inverse string tension $\alpha'$ between the holomorphic and anti-holomorphic sectors. The level-matching condition is then flipped, leading to a finite physical spectrum. Accordingly, the moduli-space integrals in their amplitudes encode the exchange of a {\it finite} set of internal states. The chiral correlators and their cohomology decompositions, however, can be freely translated between twisted and conventional strings \cite{Huang:2016bdd,Jusinskas:2019dpc}.

Because of the finite spectrum, interactions among massless and massive states of the heterotic twisted strings can be exactly described by a Lagrangian, making calculations simpler. 
According to \cite{Johansson:2017srf,Azevedo:2018dgo},
the $\alpha'\rightarrow \infty$ limit of the theory is related to a four-derivative massless supergravity that becomes conformal in four dimensions. A massless 4D Lagrangian was derived in \cite{Johansson:2018ues,Butter:2016mtk}, and this theory is equivalent to Witten's twistor string, containing both ${\cal N}=4$ SYM and conformal supergravity~\cite{Witten:2003nn,Berkovits:2004jj}.  

Here we will use a subsector of the 10D Lagrangian of the twisted string to reverse-engineer the chiral correlator for $n$-point scattering of gauge multiplets and a single massive state. This correlator can then be exported to conventional string theories to obtain the fully simplified open- 
and closed-string tree amplitudes involving a mass-level-one state,
with manifest double-copy structure and 
their exact $\ap$-dependence.
As a byproduct, the chiral $n$-point correlators also determine the
gravitational couplings of a single massive state in the twisted string.
Further details will appear in a longer paper~\cite{inprogr}.

\vspace{-0.5cm}

\section{Basics of heterotic strings}

\vspace{-0.3cm}
\noindent
We begin by reviewing the twisted heterotic
string, comparing it to the conventional heterotic string.

\medskip
{\bf A. Vertex operators:} Physical states of both twisted
and conventional closed strings are represented via vertex operators of the form
\beq
{\cal V}_{L \otimes R} = \bar V_L \otimes V_R \ e^{i k \cdot X},
\label{vops.1}
\eeq
where the polarization data factorizes into holomorphic and antiholomorphic pieces, respectively $V_R$ and $\bar V_L$. The plane
waves involve spacetime momenta $k_m$ (with
vector indices $m,n,p,\ldots = 0,1,\ldots, 9$) subject
to the mass-shell condition $k^2+ M^2=0$. We leave implicit the normal ordering with respect to the
Wick contractions
\beq
X^m (z) X^n (w) \sim - \tfrac{\ap}{2} \eta^{mn} [ \log(z{-}w) \pm \log( \bar z{-}\bar w)],
\label{vops.2}
\eeq
with signature $ \eta^{mn} = {\rm diag}(-1,1,1,\ldots,1)$. The relative $\pm$ sign is positive for conventional
and negative for twisted strings, and it propagates to the Koba--Nielsen factors
\begin{eqnarray}
\KN_{\pm} & = & \left\langle \nonumber \prod_{j=1}^n e^{i k_j \cdot X(z_j)} 
\right  \rangle_{\! \! \pm}\! \! \! , \nonumber \\ & = &  \prod_{1\leq i < j}^n 
(z_{ij})^{\tfrac{\alpha'}{2}k_i\cdot k_j} (\bar z_{ij})^{\pm  \tfrac{\alpha'}{2}k_i\cdot k_j},  \label{vops.3} 
\end{eqnarray}
with $z_{ij} = z_i{-}z_j $.

The physical spectrum of twisted heterotic strings is described by \eqref{vops.1}
with the following chiral halves in canonical superghost pictures depending on
$\varphi$:
\beq
\begin{array}{c | c | c}
\te{lv.} &\te{bosonic} \ \bar V_L&\te{supersymmetric} \ V_R
\\\hline
0&\bar V^a_{J} \sim \bar{J}^a &V_{\ep} \sim \ep_m \lamb^m e^{-\varphi}  \\
&\bar V_{\bar \epsilon} \sim \bar \epsilon_m i \bar\partial X^m_{\pm}  &V_{\chi} \sim \chi^\alpha S_\alpha e^{- \frac{\varphi}{2}} \\\hline
1 &\bar J^a \bar J^b &V_{\phi} \sim \phi_{mn} i \partial X^m_{\pm} \lamb^n e^{-\varphi} \\
&  i \bar\partial X^m_{\pm}  \bar{J}^a &V_{e} \sim e_{mnp} \lamb^m\lamb^n\lamb^p e^{-\varphi} \\
& i \bar\partial X^m_{\pm} i \bar\partial X^n_{\pm}  &V_{\psi} \sim \psi_m^\alpha
( i \partial X^m_{\pm}S_{\alpha} {-} \tfrac{\ap}{4} \! \! \not \! k_{\alpha \beta} S^{m\beta} )e^{-\frac{ \varphi}{2}}
\end{array} 
\nonumber
\eeq
The bosonic side involves Kac--Moody currents $\bar J^a$ with adjoint
indices $a,b,\ldots = 1,2, \ldots ,\te{dim}({\cal G})$ of an unspecified
gauge group ${\cal G}$ with generators $T^a$ satisfying $[T^a, T^b] = c^{abc} T^c$. The supersymmetric side contains the matter variables $\lambda^m, S_\alpha, S_m^{\beta}$ of the Ramond-Neveu-Schwarz (RNS) superstring \cite{Friedan:1985ey, Friedan:1985ge, Cohn:1986bn,Gross:1985rr}, with Weyl-spinor indices $\alpha,\beta,\ldots = 1,2,\ldots,16$.  The $SO(1,9)$ Pauli matrices satisfy $\{ \gamma^{m},\gamma^{n}\} = 2 \eta^{mn}$, and we are using $\! \not \! k_{\alpha \beta}\equiv k_m \gamma^{m}_{\alpha \beta}$.

The massless states depend on the transverse polarization vectors $\epsilon_m,\bar \epsilon_m$ 
and a chiral spinor satisfying $\not \!   k_{\alpha \beta} \chi^\beta =0$. The massive states are given by a symmetric traceless tensor $\phi_{mn}$, a 3-form $e_{mnp}$ and a $\gamma$-traceless vector-spinor $\psi_m^\alpha$
subject to $k^{m}e_{mnp}=k^{m}\phi_{mn}=k^{m}\psi_{m}^{\alpha}=0$.

The physical vertex operators are organized into three multiplets of 10D ${\cal N}=1$ supersymmetry:

\begin{adjustwidth}{0.3cm}{}
$\bullet$ a gauge multiplet involving gluon ($A$) and gluino (${\cal X})$,
\beq
{\cal V}^a_{A}  = \bar V^a_{\bar J} \otimes V_{\ep} \, e^{i k \cdot X}  \, , \ \ \ \ 
{\cal V}^a_{{\cal X}}  = \bar V^a_{\bar J} \otimes V_{\chi} \, e^{i k \cdot X} 
\label{gauge-vertex},
\eeq
$\bullet$ a supergravity multiplet involving graviton, $B$-field and dilaton ($ \bar V_{\bar \epsilon} \otimes V_{\ep}$) 
as well as gravitino and dilatino ($ \bar V_{\bar \epsilon} \otimes V_{\chi}$),

\noindent
$\bullet$ a massive multiplet with $k^2 = - \frac{4}{\ap}$ comprising a spin-$2$ field $\Phi_{mn}$,
a 3-form $E_{mnp}$ and a spin-$\tfrac{3}{2}$ field $\Psi^{\alpha}_{m}$,
\beq
{\cal V}_{\{\Phi,E,\Psi\}}  = \bar V_T \otimes V_{\{\phi,e,\psi\}} \, e^{i k \cdot X}\, .
\label{massive-vertex}
\eeq 
\end{adjustwidth}
\noindent
The massive states can be viewed as a double copy of a tachyon, $\bar V_T=1$, with 
the first mass level of the open superstring \cite{Koh:1987hm}. This construction hinges on the twisted level-matching condition.

\medskip
{\bf B. Tree-level amplitudes:} $n$-point tree-level string amplitudes are given by an integral over the moduli space
$\mathfrak{M}_{0;n}$ of $n$-punctured Riemann spheres. The integrand is the CFT correlator of $n$ string vertices, with the freedom to fix any triplet of punctures via ${\rm SL}_2(\mathbb C)$.
The conventional and twisted string amplitudes, respectively  ${\cal M}_{+}$ and ${\cal M}_{-}$, only differ in the Koba--Nielsen factor~\eqref{vops.3} and can be cast as
\begin{align}
{\cal M}_{\pm}(1,\ldots,n) &=   \int_{\mathfrak{M}_{0;n}} \! \!  \frac{ \dd^2 z_1 \ldots \dd^2 z_n}{\pi^{n-3} {\rm vol} \, {\rm SL}_2(\mathbb C)} \;
\KN_{\pm} \, \bar{{\cal I}}_L  \, {\cal I}_R .
\label{twistedstringamplitude}
\end{align}
Both explicitly factorize the main quantities of interest here: the
{\it chiral correlators}  $ {\cal I}_R$ ($\bar{{\cal I}}_L$). They are rational functions of  $z_j$ ($\bar z_j$) and multilinear in the polarizations of the chiral halves $V_R$ ($\bar V_L$),
therefore a key origin of double-copy structures.  

The integrals~\eqref{twistedstringamplitude} can be expressed in terms of the Kawai--Lewellen--Tye (KLT) formula~\cite{Kawai:1985xq, Bern:1998sv, BjerrumBohr:2010hn} as 
bilinears in disk integrals, with a sign flip of $\ap$ in one of the factors to describe ${\cal M}_{-}$  \cite{Huang:2016bdd}.
The sphere integrals in ${\cal M}_{+}$ feature an infinite number of poles for integer values of the generalized Mandelstam variables,
$s_{ij \ldots p} = \tfrac{ \ap }{4}(k_i + k_j+\ldots + k_p)^2 $. In contrast, the sphere integrals with $\KN_{-}$  evaluate to rational functions of
$s_{ij \ldots p}$ and match the pole structure of a QFT with finite mass spectrum: $M^2 =0$ and $M^2 = \frac{4}{\ap}$.

\vspace{-0.15cm}

\section{Field-theory perspective}

\vspace{-0.3cm}

\noindent
We here translate the three-point amplitudes ${\cal M}_{-}$ of one massive
vertex \eqref{massive-vertex} and two gauge multiplets into the corresponding QFT Feynman vertices. Their gauge-covariant completion 
deforms the Lagrangian of 10D ${\cal N}=1$ SYM, and the combined Feynman rules suffice to determine the chiral correlators ${\cal I}_R$ for one massive state and any number of massless ones.

\medskip
{\bf A. Three-point amplitudes:} The prescription above yields
the well-known three-point SYM amplitudes
\begin{align}
{\cal M}_-(A_1,A_2,A_3) &= 2 c^{a_1 a_2 a_3}(\ep_1{\cdot} \ep_2)(k_1{\cdot} \ep_3) + {\rm cyc}(1,2,3), \notag \\
{\cal M}_-(A_1,{\cal X}_2,{\cal X}_3) &= - c^{a_1 a_2 a_3} (\chi_2 \!\! \not \! \epsilon_1  \chi_3),
\label{vops.12}
\end{align}
while the amplitudes with one massive state are simply
\begin{align}
{\cal M}_-(A_1,A_2,\Phi_3) &= 2 \sqrt{\ap} \delta^{a_1 a_2} \phi_{3mn} f_{1\,p}^{m} f^{pn}_{2},
\notag \\
{\cal M}_-(A_1,A_2,E_3) &= i \delta^{a_1 a_2} e_{3mnp} f_1^{mn} \ep^p_{2} ,
\notag \\
{\cal M}_-({\cal X}_1,{\cal X}_2,\Phi_3) &=- \sqrt{\ap} \delta^{a_1 a_2} \phi_{3mn} (\chi_1 \gamma^m \chi_2) k_1^n,
\label{vops.13} \\
{\cal M}_-({\cal X}_1,{\cal X}_2,E_3) &=- \tfrac{i}{12} \delta^{a_1 a_2} e_{3}^{mnp}(\chi_1 \gamma_{mnp} \chi_2) ,
\notag \\
{\cal M}_-(A_1,{\cal X}_2,\Psi_3) &=i \delta^{a_1 a_2}f_1^{mn} (\chi_2 \gamma_m \psi_{3n}) ,
\notag 
\end{align}
with $f_i^{mn} = k_i^m \ep_i^n - k_i^n \ep_i^m$. Their kinematic factors are identical to those found in the 
open-superstring amplitudes at the corresponding mass levels \cite{Liu:1987tb, Anchordoqui:2008hi, Feng:2010yx}. 

\medskip
{\bf B. Lagrangian:} 

The finite spectrum of the twisted heterotic string motivates to investigate a Lagrangian description of the massive amplitudes 
\eqref{vops.13}. We will discuss three kinds of contributions
\beq
{\cal L}^-_{{\rm het}} = {\cal L}_{\rm SYM} + {\cal L}_{\rm linear}+ {\cal L}_{\rm quad} + \ldots,
\label{vops.15}
\eeq
starting with the standard Lagrangian of ${\cal N}=1$ SYM,
\begin{align}
 {\cal L}_{\rm SYM} &= {\rm Tr}\{  
 {-}\tfrac{1}{4} F_{mn} F^{mn}+ \tfrac{i}{2} ({\cal X} \gamma^m \nabla_m {\cal X})
 \}, \notag \\
 F_{mn}&= \partial_m A_n- \partial_n A_m - i [A_m,A_n],
 \label{vops.16} \\
  \nabla_m {\cal X} &=  \partial_m {\cal X} - i [ A_m , {\cal X}] \, ,
  \notag
\end{align}
where we set $g_{\rm YM}$ and the gravitational coupling $\kappa$ to 1 throughout this letter.
The second term $ {\cal L}_{\rm linear}$ in \eqref{vops.15} contains all the gauge interactions 
linear in the massive fields, and we will argue that they are exhausted by the 
gauge covariantized three-point interactions~\eqref{vops.13},
\begin{multline}
\mathcal{L}_{\text{linear}} = E_{mnp}\text{Tr}\{A^{m}\partial^{n}A^{p}-\tfrac{i}{3}A^{m}[A^{n},A^{p}]\}\\
- \tfrac{i}{24} E_{mnp}\text{Tr}(\mathcal{X}\gamma^{mnp}\mathcal{X}) 
- \sqrt{\ap} \Psi_{m}^{\alpha}\text{Tr}\{F^{mn}(\gamma_{n}\mathcal{X})_{\alpha}\} \\
  + \sqrt{\ap} \Phi_{mn}\text{Tr}\{F^{mp}F_{\phantom{n}p}^{n}-\tfrac{i}{2}(\mathcal{X}\gamma^{m}\nabla^{n}\mathcal{X})\} .
\label{interaction-lagrangian}
\end{multline}

The third term ${\cal L}_{\rm quad} $ in \eqref{vops.15} contains the kinetic terms of  the massive fields, which are not explicitly needed here. They are uniquely specified and can be mapped to a Kaluza-Klein multiplet of 11D supergravity. The ellipsis in \eqref{vops.15} features also a standard ${\cal N}=1$ supergravity sector with couplings to any combination of gauge multiplets and massive states. Moreover, we are omitting interaction terms of more than one massive state. 

A central claim of our proposal is that ${\cal L}^-_{{\rm het}}$ has no further
operators involving one massive state and an arbitrary number of gauge multiplets. As a first consistency check, we have reproduced all 
four-point and bosonic five-point string amplitudes with a single massive state
from the Lagrangian terms given here. 

A more general argument can be made to rule out higher-point interactions of the form $\sqrt{\ap} \Phi  {\rm Tr} \{ F^2 (\ap F)^N\}$. 
 Based on previous work \cite{Johansson:2017srf,Azevedo:2018dgo},  the tensionless limit $\ap \to \infty$ should be well behaved and result in a four-derivative supergravity theory that is classically conformal after dimensional reduction to 4D.
In this limit, we may redefine $\sqrt{\ap} \Phi_{mn}\rightarrow \Phi_{mn}$ to get a dimensionless and massless field that recombines with the standard graviton into the gravitational field, see e.g.~\cite{Johansson:2018ues, Azevedo:2018dgo, Berkovits:2018jvm,Azevedo:2019zbn}. But the field-strength factor $ (\ap F)^N$ cannot absorb $\ap$ since $F_{mn}$ must have dimension two in a conformal theory.  Therefore, interactions of the form $\Phi  {\rm Tr}(F^{\geq3})$ and their supersymmetric completions would obstruct a well-defined tensionless limit.

On these grounds, the ellipsis in (\ref{vops.15}) does {\it not} refer to a higher-derivative expansion of an effective Lagrangian. In conventional
string theories, in turn, the description of massive spin-two scattering through an effective action
is under investigation \cite{munichgroup}.

\medskip
{\bf C.  All-multiplicity single-trace computation:} Next we describe an efficient recursive procedure to compute tree amplitudes of $(n{-}1)$ gauge multiplets and one massive state
from the Lagrangian of the twisted heterotic string. The terms given in \eqref{vops.15} grant access to the single-trace sector
in the color decomposition
\begin{align}
&{\cal M}_-(1,2,\ldots,n{-}1,\underline{n}) = \! \! \!  \sum_{\rho \in S_{n-2}}  \! \! \! {\rm Tr}(T^{a_1} T^{a_{\rho(2)} } \ldots
T^{a_{\rho(n-1)} }) \notag \\
&\hspace{1.3cm} \times {\cal A}(1,\rho(2,3,\ldots,n{-}1)|\underline{n}) + \text{multi-trace}\, .
 \label{vops.19}
\end{align} 
Here $1,2,\ldots,n{-}1$ refers to gauge-multiplet states, and the last leg $\underline{n}$ is
taken to be massive. The color-ordered single-trace amplitudes ${\cal A}$ are cyclic in
$1,2,\ldots,n{-}1$ and only receive 
contributions from Feynman diagrams involving propagating gauge multiplets
which are completely determined by ${\cal L}_{\rm SYM}+{\cal L}_{\rm linear}$.
The omitted terms in \eqref{vops.15} only affect multi-trace contributions to \eqref{vops.19}.

Since the Lagrangian ${\cal L}^-_{{\rm het}}$ only features traces over nested commutators 
of adjoint fields, the traces in the first line of \eqref{vops.19} 
must recombine into color factors that are products of $n{-}3$ structure constants
\begin{align}
&\! \! c^{a_1 | a_2 a_3 \ldots a_{n-2} | a_{n-1}} =
c^{a_1 a_2 b} c^{ba_3 d} \ldots c^{y a_{n-3}z} c^{z a_{n-2}a_{n-1}}
 \notag \\
&~~~~~  \! \! \, = 
{\rm Tr}( [[\ldots [[ T^{a_1},T^{a_2}],T^{a_3}],\ldots ], T^{a_{n-2}}]T^{a_{n-1}})\, .
\label{defccs}
\end{align}
As a consequence, permutations of ${\cal A}(1,2,\ldots,n{-}1|\underline{n}) $ satisfy Kleiss--Kuijf amplitude relations~\cite{Kleiss:1988ne}, and \eqref{vops.19} can be more compactly written in a minimal color basis
\begin{align}
&{\cal M}_-(1,2,\ldots,n{-}1,\underline{n}) = 
 \! \! \! \sum_{\rho \in S_{n-3}} \! \! \!  
c^{a_1 | \rho(a_2  \ldots a_{n-2}) | a_{n-1}} 
 \label{vops.19A} \\
&\hspace{1.5cm} \times  {\cal A}(1,\rho(2,\ldots,n{-}2),n{-}1|\underline{n}) + \text{multi-trace}\, ,
\notag
\end{align} 
in direct analogy with the Dixon--Del Duca--Maltoni decomposition of
gauge-theory amplitudes \cite{DelDuca:1999rs}.

\medskip
{\bf D. Perturbiners:} We use the perturbiner method
\cite{Bardeen:1995gk, Rosly:1996vr, Rosly:1997ap, Selivanov:1998hn, Selivanov:1999as, Lee:2015upy, Mafra:2015vca, Mizera:2018jbh,Lopez-Arcos:2019hvg} to organize the diagrammatic computation of the color-ordered amplitudes in \eqref{vops.19}. 
To each ordered word $P = 12\ldots$ in external-particle
labels (letters), we associate multi-particle momenta $k_P = k_{1}+k_{2}+\ldots$ and multi-particle polarizations such as $\ep^m_P,f_P^{mn},\chi_P^\alpha$ which 
are identified with Berends--Giele currents~\cite{Berends:1987me}. The gauge-multiplet recursions in the Lorenz gauge
\begin{eqnarray}
\epsilon^m_{P} &=& \frac{1}{k_P^2} \sum_{P=QR} [ \ep^m_R (k_R \cdot \ep_Q) + \ep_{Rn} f^{mn}_Q   \notag \\
&& \ \ \ \ \ \ \ \ \ \ \ \ \ \ \ \ \ \ + \tfrac{1}{2}(\chi_R \gamma^m \chi_Q)  -(Q\leftrightarrow R)], \notag \\
f_P^{mn} &=& k_P^m \ep_P^n - k_P^n \ep_P^m 
- \sum_{P=QR} (\ep_Q^m \ep_R^n - \ep_Q^n \ep_R^m), \\
\chi_P^\alpha &=& \frac{(\not \! k_P)^{\alpha \beta}}{k_P^2} \sum_{P=QR} 
 [ \ep_Q^m (\gamma_m \chi_R)_\beta - (Q\leftrightarrow R) ], \notag
\end{eqnarray}
involve sums over all order-preserving deconcatenations of $P=QR$ into non-empty words $Q$ and $R$. The recursion ends with single-particle labels, defined by the on-shell polarizations.

For the massive fields,  \eqref{vops.15} leads to similar recursions
with the following single-trace contributions of gauge multiplets:
\begin{align}
\phi_{P}^{mn} &=  \sqrt{\ap} \sum_{P=QR} \big[ f_{Q \, p}^{m}f_{R}^{pn}+\tfrac{1}{2}(\chi_{Q}\gamma^{m}k_{R}^{n}\chi_{R}) \big]  \notag \\
  &\ \ - \tfrac{\sqrt{\ap}}{2}\! \!\sum_{P=QRS}  (\chi_{Q}\gamma^{m} \epsilon_{R}^n \chi_{S})+{\rm cyc}_P\,, \notag \\
e_{P}^{mnp} &=  i \sum_{P=QR}\big[ \epsilon_{Q}^{m}k_{R}^{n}\epsilon_{R}^{p}-\tfrac{1}{24}(\chi_{Q}\gamma^{mnp}\chi_{R})\big]   \label{bgmassive} \\
  &\ \ - \tfrac{2i}{3}\! \!\sum_{P=QRS} \epsilon_{Q}^{m}\epsilon_{R}^{n}\epsilon_{S}^{p}+{\rm cyc}_P \, , 
\notag\\
\bar{\psi}_{P\alpha}^{m} &=  - i \sum_{P=QR} f_{Q}^{mp}(\gamma_{p}\chi_{R})_{\alpha}+{\rm cyc}_P \, . \notag
\end{align}
The notation $+{\rm cyc}_P$ instructs to add cyclic permutations of the letters in $P$. In this way, the $n$-point amplitudes 
\begin{align}
{\cal A}(1,2,\ldots,n{-}1|\Phi_{n}) & = (\phi_{12 \ldots n-1})^{mp}(\phi_{n})_{mp}, \notag\\
{\cal A}(1,2,\ldots,n{-}1|E_{n}) & = (e_{12 \ldots n-1})^{mpq}(e_{n})_{mpq},  \label{perturbiner-npt}\\
{\cal A}(1,2,\ldots,n{-}1|\Psi_{n}) & = (\bar{\psi}_{12 \ldots n-1})_{\alpha}^{m}(\psi_{n})_{m}^{\alpha}, \notag
\end{align}
are cyclically invariant in $1,2,\ldots,n{-}1$.


\section{String amplitudes from QFT}

\vspace{-0.3cm}

\noindent
The introduced QFT description implies new results for a variety of string amplitudes that we now describe.

\medskip
{\bf A. Cohomology decomposition:} Consider on-shell momenta $k_j^2 = 0$ for legs $1,2,\ldots,n{-}1$ and
$k_n^2 =- \frac{4}{\ap}$ subject to $\sum_{i=1}^n k_i=0$. For this kinematic
configuration the ${\rm SL}_2(\mathbb C)$-fixing $(z_1,z_{n-1},z_n) \rightarrow (0,1,\infty)$
yields the following Koba--Nielsen factors \eqref{vops.3} and chiral correlators:
\begin{align}
 \KN_{\pm}^{\infty} &= 
\lim_{z_n \rightarrow \infty} \frac{ \KN_{\pm}  }{z_n^2 \bar z_n^{\pm 2}}
= \prod_{1\leq i < j}^{n-1} 
(z_{ij})^{s_{ij}} (\bar z_{ij})^{\pm s_{ij}} 
\label{cohom.2} \\
{\cal I}_R^{\infty} &= \lim_{z_n \rightarrow \infty} z_n^4 {\cal I}_R  \, , \ \ \ \ \ \ 
\overline{{\cal I}}_L^{\infty} = \lim_{\bar z_n \rightarrow \infty} \bar z_n^{2\pm 2} \overline{{\cal I}}_L \, .
\notag
\end{align}
The amplitude prescription \eqref{twistedstringamplitude} then specializes to
\begin{align}
{\cal M}_{\pm}(1,\ldots,n) &=   \int_{\mathbb C^{n-3}} \! \! \! \! \! \! \! \! \!   \frac{ \dd^2 z_2 \ldots \dd^2 z_{n-2}}{\pi^{n-3}} \;
\KN^{\infty}_{\pm} \, \overline{{\cal I}}^{\infty}_L  \, {\cal I}^{\infty}_R \, .
\label{vops.10B}
\end{align}
For $n{-}1$ gauge multiplets, the  single-trace contributions  to the antiholomorphic chiral correlators are \cite{Frenkel1992}
\begin{align}
\overline{{\cal I}}_L^{\infty} &= \langle \bar J^{a_1}(z_1)  \bar J^{a_2}(z_2) \ldots \bar J^{a_{n-1}}(z_{n-1}) \rangle_{\te{single trace}}
\label{jbasis} \\
&= \! \! \!  \! \sum_{\rho \in S_{n-3}} \! \! \!  \! c^{a_1 | \rho(a_2 a_3 \ldots a_{n-2}) | a_{n-1}} \overline{ {\rm PT}(1,\rho(2,\! \ldots \! ,n{-}2),n{-}1)},
\notag
\end{align}
where $c^{a_1 | a_2 a_3 \ldots a_{n-2} | a_{n-1}} $ denote color factors \eqref{defccs}, and the open chains 
\beq
{\rm PT}(1,2,\ldots ,n{-}1) = \frac{1}{z_{12} z_{23}\ldots z_{n-3,n-2} z_{n-2,n-1}}
\label{cohom.4}
\eeq
arise from ${\rm SL}_2(\mathbb C)$-fixed Parke--Taylor factors. The permutations of ${\rm PT}$ 
with $\rho \in S_{n-3}$ in \eqref{jbasis} form a basis of chiral integrands under integration by 
parts, i.e.\ they span the twisted cohomology defined by the Koba--Nielsen factor \eqref{vops.3} \cite{Aomoto}. Accordingly, the holomorphic chiral correlator can be expressed in a Parke--Taylor basis as
\beq
 \! \! \! {\cal I}_R^{\infty} = \!  \! \! \!  \sum_{\rho \in S_{n-3}}   \!  \!  \! \! 
K^{1 | \rho(2  \ldots n{-}2) |n{-}1}  
\, {\rm PT}(1,\rho(2,\ldots  ,n{-}2),n{-}1).
\! \label{cohom.5}
\eeq
The kinematic factors $K^{1|\rho|n-1}$ are multilinear in the polarizations of
$n{-}1$ gauge multiplets and the massive state.

\medskip
{\bf B. The supersymmetric chiral correlator:} The kinematic factors in \eqref{cohom.5} can be found via a matching with the QFT computation described above. This is possible because 
the $(n{-}3)! \times (n{-}3)!$ matrix of Parke--Taylor integrals in \eqref{vops.10B}
is non-degenerate, i.e.\ there is a matrix inverse $S_0 = m^{-1}$ of
\begin{align}
&m(\rho|\sigma) = 
 \int_{\mathbb C^{n-3}} \! \! \! \! \! \! \! \! \!   \frac{ \dd^2 z_2 \ldots \dd^2 z_{n-2}}{\pi^{n-3}} \; \KN^{\infty}_{-}
 \label{cohom.31}\\
 &\times {\rm PT}(1,\rho(2,\ldots,n{-}2),n{-}1) \, \overline{  {\rm PT}(1,\sigma(2,\ldots,n{-}2),n{-}1) }. \,
 \notag
\end{align}
For instance, $S_0(\emptyset | \emptyset) = 1$ at $n=3$ and $S_0(2 |2) = \frac{ s_{12} s_{23} }{s_{12}{+}s_{23}}$
at $n=4$. In fact, $m(\rho|\sigma)$ and its inverse $S_0(\rho|\sigma)$ coincide with the doubly-partial amplitudes
of bi-adjoint scalars \cite{Cachazo:2013iea} and the KLT kernel \cite{Kawai:1985xq, Bern:1998sv, BjerrumBohr:2010hn} for massless external states as long as all expressions are
written in terms of the kinematic variables
\beq
{\bf s}_n = \{ s_{ij} , \ 1\leq i<j\leq n{-}1\} \setminus \{ s_{1,n-1} \}
\label{cohom.32}
\eeq
that occur in $\KN_{\pm}^{\infty}$,
where $s_{1,n-1}$ drops out by
$z_{n-1,1}=1$ in our ${\rm SL}_2$-fixing. 
However, the relation $\sum_{1\leq i<j}^{n-1} s_{ij}= \frac{\ap}{4} k_n^2$ to reinstate
$s_{1,n-1}$ varies with the mass of the $n^{\rm th}$ leg.

Using the chiral correlator representations \eqref{jbasis}-\eqref{cohom.5}, and the
sphere integrals \eqref{cohom.31}, the kinematic factors $K^{1|\rho|n-1}$
in ${\cal I}_R^{\infty} $ are uniquely determined to be
\begin{align}
{\cal I}_R^{\infty} &= \!  \sum_{\rho,\sigma \in S_{n-3}} \! 
{\rm PT}(1,\rho(2,\ldots,n{-}2),n{-}1)
\label{cohom.33} \\
&\times S_0(\rho|\sigma) {\cal A}(1,\sigma(2,\ldots,n{-}2),n{-}1|\underline{n}) ,
\notag
\end{align}
by requiring the color-dressed amplitude (\ref{vops.19A}) to be reproduced by
the twisted heterotic string. This chiral correlator and the QFT construction of
${\cal A}$ using perturbiner techniques are the main result of this letter. Their value will be underlined by the subsequent applications.

\medskip
{\bf C. Implications for other twisted heterotic string amplitudes:} Using the above ingredients, it is straightforward to compute multi-trace or gravitational amplitudes.
For example, the four-point amplitude of two gauge multiplets $1,3$, a 
gravitational multiplet $2_h$ and a massive multiplet $\underline{4}$ follows
from \eqref{vops.10B} along with ${\cal I}_R^{\infty}$ in \eqref{cohom.33} and
\beq
\overline{{\cal I}}_L^{\infty} = -2\ap \delta^{a_1 a_3} \bigg(
 \frac{ \bar \epsilon_2\cdot k_1 }{\bar z_{21}} 
 + \frac{ \bar \epsilon_2\cdot k_3 }{\bar z_{23}} \bigg)
 \label{cohom.36}
\eeq
The sphere integrations then yield
\beq \label{cohom.37} 
{\cal M}_-(1,2_h, 3,\underline{4}) \!=\! \frac{2\ap \! \delta^{a_1\! a_3}}{1+s_{13}}  {\cal A}(1,  2, 3|\underline{4})\,
 (s_{23}\bar \epsilon_2{\cdot} k_1 - s_{12}\bar \epsilon_2{\cdot }k_3)
\eeq
which exhibits the expected massless poles from gauge-multiplet exchange
in the $s_{12},s_{23}$ channels and poles in $s_{13}$ and $1{+}s_{13}$
from graviton- and massive-state exchange.
The same techniques lead to all-multiplicity results for multi-trace
and gravitational amplitudes with a single massive state:
The underlying $\overline{{\cal I}}_L^{\infty} $ are straightforward
to obtain from Wick contractions of $\bar J^{a_i}(z_i)$ \& $\bar \epsilon_j{\cdot} \partial_{\bar z} X_{-}(z_j)$,
and their Parke--Taylor decompositions are well-known from conventional strings \cite{Schlotterer:2016cxa, He:2018pol, He:2019drm}.

\medskip
{\bf D. Implications for type-I superstrings:} We can also export our method to conventional strings. Tree-level amplitudes of the open 
type-I superstring with only one massive mutiplet $\underline{n}$  boil down to ${\cal I}_R \KN^{1/2}_+$ integrated over a disk boundary,
\begin{align}
&{\cal A}_{\te{type I}}(1,2,\ldots,n{-}1,\underline{n}) = \sum_{\rho \in S_{n-3}} F^{\rho}({\bf s}_n)
\label{cohom.38} \\
&\ \ \ \ \ \ \ \ \ \  \times  {\cal A}(1,\rho(2,\ldots,n{-}2),n{-}1|\underline{n}) \big|_{\alpha' \rightarrow 4 \alpha'}  \, ,
\notag
\end{align}
where we have fixed $z_1=0$ and $z_{n-1}=1$ in
\begin{align}
&F^{\rho}({\bf s}_n) =  \! \!  \! \! \! \! \! \! \! \! \! \!   \int \limits_{0<z_2<z_3<\ldots <z_{n-2}<1}  \! \! \! \!    \! \! \! \! \! \! \! \! \! \! \! \! 
\dd z_2 \,\dd z_3\ldots \dd z_{n-2} \prod_{1\leq i<j}^{n-1} |z_{ij}|^{s_{ij}}  \label{cohom.39}\\
&\, \times  \rho \bigg\{ \frac{s_{21} }{z_{21}} \bigg( \frac{s_{31} }{z_{31}} {+}\frac{s_{32} }{z_{32}} \bigg) \ldots \bigg(\frac{s_{n-2,1} }{z_{n-2,1}} {+}\ldots {+}\frac{s_{n-2,n-3} }{z_{n-2,n-3}}  \bigg) \bigg\}\, .
\notag
\end{align}
The rescaling $\alpha' \rightarrow 4 \alpha'$ characteristic to open strings applies to the entire right-hand side of (\ref{cohom.38}).
In this case, $\underline{n}$ is
Lie-algebra valued, i.e.\ \eqref{cohom.38} is the coefficient of the $n$-trace ${\rm Tr}(T^{a_1} T^{a_2} \ldots T^{a_n})$. 
The open-string incarnation of the massive spin-2 field has been related to conformal 
supergravity in the massless limit \cite{Ferrara:2018wlb}.

As functions of the $\frac{n}{2}(n{-}3)$ Mandelstam invariants ${\bf s}_n$ in \eqref{cohom.32},
the disk integrals $F^{\rho}$ in \eqref{cohom.39} coincide with the basis in the massless 
open-string amplitudes of \cite{Mafra:2011nv}.
However, the relations between ${\bf s}_n$ and $s_{1,n-1}$ or $s_{jn}$ with $j=1,2,\ldots,n{-}1$
depend on the external masses,
i.e.\ the denominator of the four-point example \cite{Liu:1987tb, Feng:2010yx}
\beq
{\cal A}_{\te{type I}}(1,2,3,\underline{4}) = \frac{\Gamma(1{+}s_{12}) \Gamma(1{+}s_{23}) }{\Gamma(1{+}s_{12}{+}s_{23})}
 {\cal A}(1,2,3|\underline{4})
 \label{typeoneex}
\eeq
equals $\Gamma(-s_{13})$ in the massive case rather than $\Gamma(1{-}s_{13})$ as in the massless one.
Also the five-point instance of \eqref{cohom.38} for external states $\Phi_5$ \cite{svensson} or $E_5$ and four gluons has been verified via explicit integral reduction in the chiral correlator which also crosschecks fermionic component amplitudes via supersymmetry \cite{Koh:1987hm, Feng:2012bb}.

\medskip
{\bf E. Implications for type-II superstrings:} Similarly, the sphere integral ${\cal M}_{+}$ in \eqref{vops.10B} with
${\cal I}_L^{\infty}={\cal I}_R^{\infty}$ determines type-II amplitudes
with a single mass-level-one multiplet from two copies of \eqref{cohom.33},
\begin{align}
&{\cal M}_{\te{type II}}(1,2,\ldots,n{-}1,\underline{n}) = \sum_{\rho,\sigma \in S_{n-3}} G_{\rho|\sigma}({\bf s}_n)
\label{cohom.40} \\
& \ \times \! {\cal A}(1,\rho(2,\! \ldots \!,n{-}2),n{-}1|\underline{n}) \overline{ {\cal A}(1,\sigma(2,\! \ldots \!,n{-}2),n{-}1|\underline{n})}\, .
\notag
\end{align}
The double-copy of the states $\{ \Phi,E,\Psi\}$ in \eqref{vops.15} does not arise
in the twisted heterotic string and involves spins ranging between 0 and 4.
As functions of the Mandelstam basis in \eqref{cohom.32}, the
sphere integrals $ G_{\rho|\sigma}({\bf s}_n)$ coincide with those in
the expansion of massless type-II amplitudes in terms of ${\cal A}_{\rm SYM}(1,\rho,n{-}1,n)
\overline{{\cal A}_{\rm SYM}(1,\sigma,n{-}1,n)}$ \cite{Schlotterer:2012ny}. 

In the four-point example
\begin{align}
G_{2|2}({\bf s}_4) &=  \frac{1 }{\pi}
 \bigg( \frac{s_{12}s_{23}}{s_{12}{+}s_{23}} \bigg)^2 \! \int \limits_{\mathbb C}
\dd^2 z_2 \,   | z_{2}|^{2s_{12}-2} | 1{-}z_2|^{2s_{23}-2}
\notag \\
&= s_{12}s_{13}s_{23}   \frac{ \Gamma(1{+}s_{12})\Gamma(1{+}s_{13}) \Gamma(1{+}s_{23}) 
}{\Gamma(1{-}s_{12}) \Gamma(1{-}s_{13}) \Gamma(1{-}s_{23})}\, ,
\label{cohom.41}
\end{align}
the prefactor of $ s_{12}s_{13}s_{23}$ cancels the massless double poles from
${\cal A}(1,2,3|\underline{4}) \overline{{\cal A}(1,2,3|\underline{4})}$. It is striking that the QFT
computation of the kinematic factors ${\cal A}$ completely fixes the polarization dependence
of the string amplitudes \eqref{cohom.38} and \eqref{cohom.40} where the propagation of the complete
massive spectrum is reflected by well-studied scalar integrals $F^\rho$ and $G_{\rho|\sigma}$.
\vspace{-0.15cm}
\section{Conclusions and further directions}

\vspace{-0.3cm}
\noindent
We have developed here a new method combining QFT and string-theory techniques to obtain all-multiplicity tree amplitudes (exact in $\alpha'$) with a massive external state. Our results readily apply to the
gauge and gravity sectors of the twisted heterotic string as well as type-I and type-II superstrings.

The backbone of our construction is a cohomology decomposition of the moduli-space integrands,
which is known~\cite{Aomoto} to directly generalize to string amplitudes with several massive states, as well as higher mass levels of conventional string theories. It is remarkable that the currently known all-multiplicity coefficients have a QFT interpretation for every type of string theory. For several massive states, or higher mass-level states (see for instance \cite{Bianchi:2010es, Feng:2011qc} for detailed studies of mass level 2), one may expect that the coefficients of the cohomology decompositions continue to exhibit structural simplicity, which hopefully stems from a QFT perspective. Since the twisted heterotic string does not admit higher mass-level states, such a construction goes beyond the scope of the current treatment. 

A more direct generalization is to formulate the obtained Lagrangian and amplitudes in pure-spinor superspace, based on massive vertex operators \cite{Berkovits:2002qx, Chakrabarti:2018mqd, Chakrabarti:2018bah}. Besides manifest spacetime supersymmetry, this gives access to BRST-cohomology methods.  Similarly, we expect our techniques to be useful at loop level: The Lagrangian description of massive states may
shed light on the open questions on loop amplitudes of twisted strings, and feed into conventional-string amplitudes, for instance, via unitarity cuts.

Potential physics applications of massive string amplitudes include exploring chaos in the scattering
of excited string states \cite{Gross:2021gsj}, motivated by their correspondence with black-hole microstates \cite{Horowitz:1996nw}. The relevance of excited string states for black-hole physics, causality and unitarity led to a regained interest in their scattering amplitudes \cite{DAppollonio:2015fly, Skliros:2016fqs, Bianchi:2019ywd}. Moreover, massive strings resonances may become relevant at colliders in the case of a low string scale \cite{ArkaniHamed:1998rs, Antoniadis:1998ig, Feng:2010yx}.

An interesting generalization of our work is to consider amplitudes with two massive modes: Such higher-spin massive amplitudes were recently used for describing classical Kerr black-hole scattering~\cite{Guevara:2018wpp}, needed for binary inspiral and gravitational wave physics. Similarly, there has been a revival of string-amplitude methods for black-hole eikonal scattering~\cite{Amati:1988tn, DiVecchia:2020ymx}, which can benefit from better knowledge of massive string amplitudes. Finally, massive string amplitudes in flat spacetime also carry relevant information for the AdS/CFT correspondence, as for instance showcased in \cite{Bargheer:2013faa,Minahan:2014usa,Antunes:2020pof}.


\vspace{0.4cm}

\begin{acknowledgments}
{\bf Acknowledgments:} We are grateful to Anders Svensson and Fei Teng for inspiring discussions on related work.
Moreover, Dieter L\"ust and Tomasz Taylor are thanked for valuable comments on a draft.
MG and OS are supported by the European Research Council under ERC-STG-804286 UNISCAMP.  
HJ is supported by the Knut and Alice Wallenberg Foundation under grants KAW 2018.0116, KAW 2018.0162, and the Ragnar S\"{o}derberg Foundation (Swedish Foundations' Starting Grant).
RLJ  acknowledges the support from ESIF and MEYS through the project 
CZ.02.2.69/0.0/0.0/18\_053/0016627.
\end{acknowledgments}



\vspace{-0.5cm}

\vskip .3 cm

\begin{thebibliography}{99}
\vspace{-0.5cm}

\bibitem{Kawai:1985xq}
H.~Kawai, D.~C.~Lewellen and S.~H.~H.~Tye,
{\emph{Nucl.\ Phys.\ B} \textbf{269} (1986), 1-23}


\bibitem{Bern:2008qj}
Z.~Bern, J.~J.~M.~Carrasco and H.~Johansson,
{\emph{Phys.\ Rev.\ D} \textbf{78} (2008) 085011},
[\href{http://arxiv.org/abs/0805.3993}{{\tt 0805.3993}}].

\bibitem{Bern:2019prr}
Z.~Bern, J.~J.~Carrasco, M.~Chiodaroli, H.~Johansson and R.~Roiban,
[\href{http://arxiv.org/abs/1909.01358}{{\tt 1909.01358}}].

\bibitem{chisplitt}
E.~D'Hoker and D.~H.~Phong, {\emph{Commun.\ Math.\ Phys.}\ \textbf{125} (1989) 469}.

\bibitem{Bern:2010ue}
Z.~Bern, J.~J.~M.~Carrasco and H.~Johansson,
{\emph{Phys.\ Rev.\ Lett.}\ \textbf{105} (2010) 061602},
[\href{http://arxiv.org/abs/1004.0476}{{\tt 1004.0476}}].

\bibitem{Green:1987sp}
M.~B.~Green, J.~H.~Schwarz and E.~Witten,
Cambridge Univ. Pr. (1987) 469 P.

\bibitem{Polchinski:1998rq}
J.~Polchinski, 
Cambridge Univ. Pr. (1998) 424 P.

\bibitem{Mafra:2011nv}
C.~R.~Mafra, O.~Schlotterer and S.~Stieberger,
{\emph{Nucl.\ Phys.\ B}
  {\bf 873} (2013) 419}, 
  [\href{http://arxiv.org/abs/1106.2645}{{\tt 1106.2645}}]. 

\bibitem{Zfunctions}
J.~Broedel, O.~Schlotterer and S.~Stieberger, {\emph{Fortsch.\ Phys.}\ {\bf
  61} (2013) 812}, [\href{http://arxiv.org/abs/1304.7267}{{\tt 1304.7267}}].

\bibitem{Huang:2016tag}
Y.~t.~Huang, O.~Schlotterer and C.~Wen,
{\emph{JHEP} \textbf{09} (2016) 155}, [\href{http://arxiv.org/abs/1602.01674}{{\tt 1602.01674}}]

\bibitem{Azevedo:2018dgo}
T.~Azevedo, M.~Chiodaroli, H.~Johansson and O.~Schlotterer,
 {\emph{JHEP} {\bf 10} (2018) 012}, [\href{http://arxiv.org/abs/1803.05452}{{\tt 1803.05452}}].
  
\bibitem{Mizera:2017cqs}
S.~Mizera,
{\emph{JHEP} \textbf{08} (2017) 097}
[\href{http://arxiv.org/abs/1706.08527}{{\tt 1706.08527}}].

\bibitem{Mizera:2017rqa}
S.~Mizera,
{\emph{Phys.\ Rev.\ Lett.} \textbf{120} (2018) no.14, 141602}
[\href{http://arxiv.org/abs/1711.00469}{{\tt 1711.00469}}].
  
\bibitem{Schlotterer:2012ny}
  O.~Schlotterer and S.~Stieberger,
 {\emph{J.\ Phys.\ A} {\bf 46} (2013) 475401}, [\href{http://arxiv.org/abs/1205.1516}{{\tt 1205.1516}}].  
  
\bibitem{Stieberger:2013wea}
S.~Stieberger,
{\emph{J.\ Phys.\ A} \textbf{47} (2014) 155401},
[\href{http://arxiv.org/abs/1310.3259}{{\tt 1310.3259}}].
  
\bibitem{Stieberger:2014hba}
S.~Stieberger and T.~R.~Taylor,
{\emph{Nucl.\ Phys.\ B} \textbf{881} (2014) 269-287},
[\href{http://arxiv.org/abs/1401.1218}{{\tt 1401.1218}}].
  

  
  


\bibitem{Hohm:2013jaa}
O.~Hohm, W.~Siegel and B.~Zwiebach,
{\emph{JHEP} \textbf{02} (2014) 065}, [\href{http://arxiv.org/abs/1306.2970}{{\tt 1306.2970}}]

\bibitem{Huang:2016bdd}
Y.~t.~Huang, W.~Siegel and E.~Y.~Yuan,
{\emph{JHEP} \textbf{09} (2016) 101},
[\href{http://arxiv.org/abs/1603.02588}{{\tt 1603.02588}}].


\bibitem{Jusinskas:2019dpc}
R.~Lipinski Jusinskas,
{\emph{JHEP} \textbf{12} (2019) 143},
[\href{http://arxiv.org/abs/1909.04069}{{\tt 1909.04069}}].

\bibitem{Johansson:2017srf}
H.~Johansson and J.~Nohle,
[\href{http://arxiv.org/abs/1707.02965}{{\tt 1707.02965}}].


\bibitem{Johansson:2018ues}
H.~Johansson, G.~Mogull and F.~Teng,
 {\emph{JHEP} \textbf{09} (2018) 080},
[\href{http://arxiv.org/abs/1806.05124}{{\tt 1806.05124}}].


\bibitem{Butter:2016mtk}
D.~Butter, F.~Ciceri, B.~de Wit and B.~Sahoo,
 {\emph{Phys.\ Rev.\ Lett.} \textbf{118}, no.8 (2017) 081602},
[\href{http://arxiv.org/abs/1609.09083}{{\tt 1609.09083}}].



  
  \bibitem{Witten:2003nn}
E.~Witten,
 {\emph{Commun.\ Math.\ Phys.} \textbf{252} (2004) 189-258},
[\href{https://arxiv.org/abs/hep-th/0312171}{{\tt hep-th/0312171}}].

\bibitem{Berkovits:2004jj}
N.~Berkovits and E.~Witten,
{\emph{JHEP} \textbf{08} (2004) 009},
[\href{https://arxiv.org/abs/hep-th/0406051}{{\tt hep-th/0406051}}].


\bibitem{inprogr}
M.~Guillen, H.~Johansson, R.~Jusinskas and O.~Schlotterer, work in progress.

\bibitem{Friedan:1985ey}
D.~Friedan, S.~H.~Shenker and E.~J.~Martinec,
{\emph{Phys.\ Lett.\ B} \textbf{160} (1985) 55-61}.

\bibitem{Friedan:1985ge}
D.~Friedan, E.~J.~Martinec and S.~H.~Shenker,
{\emph{Nucl.\ Phys.\ B} \textbf{271} (1986) 93-165}.

\bibitem{Cohn:1986bn}
J.~Cohn, D.~Friedan, Z.~a.~Qiu and S.~H.~Shenker,
{\emph{Nucl.\ Phys.\ B} \textbf{278} (1986) 577-600}.

\bibitem{Gross:1985rr}
D.~J.~Gross, J.~A.~Harvey, E.~J.~Martinec and R.~Rohm,
{\emph{Nucl.\ Phys.\ B} \textbf{267} (1986) 75-124.}

\bibitem{Koh:1987hm}
I.~G.~Koh, W.~Troost and A.~Van Proeyen,
{\emph{Nucl.\ Phys.\ B} \textbf{292} (1987) 201-221}.
  
\bibitem{Bern:1998sv}
Z.~Bern, L.~J.~Dixon, M.~Perelstein and J.~S.~Rozowsky,
{\emph{Nucl.\ Phys.\ B} \textbf{546} (1999) 423-479},
[\href{https://arxiv.org/abs/hep-th/9811140}{{\tt hep-th/9811140}}].

\bibitem{BjerrumBohr:2010hn}
N.~E.~J.~Bjerrum-Bohr, P.~H.~Damgaard, T.~Sondergaard and P.~Vanhove,
{\emph{JHEP} \textbf{01} (2011) 001},
[\href{http://arxiv.org/abs/1010.3933}{{\tt 1010.3933}}].

\bibitem{Liu:1987tb}
F.~Liu,
{\emph{Phys.\ Rev.\ D} \textbf{38} (1988) 1334 }.

\bibitem{Anchordoqui:2008hi}
L.~A.~Anchordoqui, H.~Goldberg and T.~R.~Taylor,
{\emph{Phys.\ Lett.\ B} \textbf{668} (2008) 373-377},
[\href{http://arxiv.org/abs/0806.3420}{{\tt 0806.3420}}].

\bibitem{Feng:2010yx}
W.~Z.~Feng, D.~L\"ust, O.~Schlotterer, S.~Stieberger and T.~R.~Taylor,
{\emph{Nucl.\ Phys.\ B} \textbf{843} (2011) 570-601},
[\href{http://arxiv.org/abs/1007.5254}{{\tt 1007.5254}}].


\bibitem{Berkovits:2018jvm}
N.~Berkovits and M.~Lize,
{\emph{JHEP} \textbf{09} (2018), 097},
[\href{http://arxiv.org/abs/1807.07661}{{\tt 1807.07661}}].

\bibitem{Azevedo:2019zbn}
T.~Azevedo, R.~L.~Jusinskas and M.~Lize,
{\emph{JHEP} \textbf{01} (2020), 082},
[\href{http://arxiv.org/abs/1908.11371}{{\tt 1908.11371}}].

\bibitem{munichgroup}
D.~L\"ust, C.~Markou, P.~Mazloumi and S.~Stieberger, work in progress

\bibitem{Kleiss:1988ne}
R.~Kleiss and H.~Kuijf,
{\emph{Nucl.\ Phys.\ B} \textbf{312} (1989) 616-644}.


\bibitem{DelDuca:1999rs}
V.~Del Duca, L.~J.~Dixon and F.~Maltoni,
{\emph{Nucl.\ Phys.\ B} \textbf{571} (2000) 51-70},
[\href{https://arxiv.org/abs/hep-ph/9910563}{{\tt hep-ph/9910563}}].


\bibitem{Bardeen:1995gk}
W.~A.~Bardeen,
{\emph{Prog.\ Theor.\ Phys.\ Suppl.} \textbf{123} (1996) 1-8}

\bibitem{Rosly:1996vr}
A.~A.~Rosly and K.~G.~Selivanov,
{\emph{Phys.\ Lett.\ B} \textbf{399} (1997) 135-140},
[\href{https://arxiv.org/abs/hep-th/9611101}{{\tt hep-th/9611101}}].



\bibitem{Rosly:1997ap}
A.~A.~Rosly and K.~G.~Selivanov,
[\href{https://arxiv.org/abs/hep-th/9710196}{{\tt hep-th/9710196}}].


\bibitem{Selivanov:1998hn}
K.~G.~Selivanov,
{\emph{Commun.\ Math.\ Phys.} \textbf{208} (2000) 671-687},
[\href{https://arxiv.org/abs/hep-th/9809046}{{\tt hep-th/9809046}}].


\bibitem{Selivanov:1999as}
K.~G.~Selivanov,
[\href{https://arxiv.org/abs/hep-th/9905128}{{\tt hep-th/9905128}}].



\bibitem{Lee:2015upy}
S.~Lee, C.~R.~Mafra and O.~Schlotterer,
{\emph{JHEP} \textbf{03} (2016) 090},
[\href{http://arxiv.org/abs/1510.08843}{{\tt 1510.08843}}].

\bibitem{Mafra:2015vca}
C.~R.~Mafra and O.~Schlotterer,
{\emph{JHEP} \textbf{03} (2016) 097},
[\href{http://arxiv.org/abs/1510.08846}{{\tt 1510.08846}}].

\bibitem{Mizera:2018jbh}
S.~Mizera and B.~Skrzypek,
{\emph{JHEP} \textbf{10} (2018) 018},
[\href{http://arxiv.org/abs/1809.02096}{{\tt 1809.02096}}].

\bibitem{Lopez-Arcos:2019hvg}
C.~Lopez-Arcos and A.~Q.~V\'elez,
{\emph{JHEP} \textbf{11} (2019) 010},
[\href{http://arxiv.org/abs/1907.12154}{{\tt 1907.12154}}].

\bibitem{Berends:1987me}
F.~A.~Berends and W.~T.~Giele,
{\emph{Nucl.\ Phys.\ B} \textbf{306} (1988) 759-808}.

\bibitem{Frenkel1992}
I.~Frenkel and M.~Zhu, 
{\emph{Duke Math J.} \textbf{66} (1992) 123}. 

\bibitem{Aomoto}
K.~Aomoto, 
{\emph{J.\ Math.\ Soc.\ Japan} \textbf{39} (1987) 191-208}.


\bibitem{Cachazo:2013iea}
F.~Cachazo, S.~He and E.~Y.~Yuan,
{\emph{JHEP} \textbf{07} (2014) 033},
[\href{http://arxiv.org/abs/1309.0885}{{\tt 1309.0885}}].


\bibitem{Schlotterer:2016cxa}
O.~Schlotterer,
{\emph{JHEP} \textbf{11} (2016) 074},
[\href{http://arxiv.org/abs/1608.00130}{{\tt 1608.00130}}].

\bibitem{He:2018pol}
S.~He, F.~Teng and Y.~Zhang,
{\emph{Phys.\ Rev.\ Lett.} \textbf{122}, no.21 (2019)  211603},
[\href{http://arxiv.org/abs/1812.03369}{{\tt 1812.03369}}].


\bibitem{He:2019drm}
S.~He, F.~Teng and Y.~Zhang,
{\emph{JHEP} \textbf{09} (2019) 085},
[\href{http://arxiv.org/abs/1907.06041}{{\tt 1907.06041}}].

\bibitem{Ferrara:2018wlb}
S.~Ferrara, A.~Kehagias and D.~L\"ust,
{\emph{JHEP} \textbf{05} (2019) 100},
[\href{http://arxiv.org/abs/1810.08147}{{\tt 1810.08147}}].


\bibitem{svensson}
A.~Svensson, unpublished.

\bibitem{Feng:2012bb}
W.~Z.~Feng, D.~L\"ust and O.~Schlotterer,
{\emph{Nucl.\ Phys.\ B} \textbf{861} (2012) 175-235},
[\href{http://arxiv.org/abs/1202.4466}{{\tt 1202.4466}}].

\bibitem{Bianchi:2010es}
M.~Bianchi, L.~Lopez and R.~Richter,
{\emph{JHEP} \textbf{03} (2011) 051},
[\href{http://arxiv.org/abs/1010.1177}{{\tt 1010.1177}}].

\bibitem{Feng:2011qc}
W.~Z.~Feng and T.~R.~Taylor,
{\emph{Nucl.\ Phys.\ B} \textbf{856} (2012) 247-277},
[\href{http://arxiv.org/abs/1110.1087}{{\tt 1110.1087}}].


\bibitem{Berkovits:2002qx}
N.~Berkovits and O.~Chandia,
{\emph{JHEP} \textbf{08} (2002) 040},
[\href{https://arxiv.org/abs/hep-th/0204121}{{\tt hep-th/0204121}}].


\bibitem{Chakrabarti:2018mqd}
S.~Chakrabarti, S.~P.~Kashyap and M.~Verma,
{\emph{JHEP} \textbf{10} (2018) 147},
[\href{http://arxiv.org/abs/1802.04486}{{\tt 1802.04486}}].

\bibitem{Chakrabarti:2018bah}
S.~Chakrabarti, S.~P.~Kashyap and M.~Verma,
{\emph{JHEP} \textbf{12} (2018) 071},
[\href{http://arxiv.org/abs/1808.08735}{{\tt 1808.08735}}].


\bibitem{Gross:2021gsj}
D.~J.~Gross and V.~Rosenhaus,
[\href{http://arxiv.org/abs/2103.15301}{{\tt 2103.15301}}].

\bibitem{Horowitz:1996nw}
G.~T.~Horowitz and J.~Polchinski,
{\emph{Phys.\ Rev.\ D} \textbf{55} (1997) 6189-6197},
[\href{https://arxiv.org/abs/hep-th/9612146}{{\tt hep-th/9612146}}].

\bibitem{DAppollonio:2015fly}
G.~D'Appollonio, P.~Di Vecchia, R.~Russo and G.~Veneziano,
{\emph{JHEP} \textbf{05} (2015) 144},
[\href{http://arxiv.org/abs/1502.01254}{{\tt 1502.01254}}].

\bibitem{Skliros:2016fqs}
D.~P.~Skliros, E.~J.~Copeland and P.~M.~Saffin,
{\emph{Nucl.\ Phys.\ B} \textbf{916} (2017) 143-207},
[\href{http://arxiv.org/abs/1611.06498}{{\tt 1611.06498}}].

\bibitem{Bianchi:2019ywd}
M.~Bianchi and M.~Firrotta,
{\emph{Nucl.\ Phys.\ B} \textbf{952} (2020) 114943},
[\href{http://arxiv.org/abs/1902.07016}{{\tt 1902.07016}}].



\bibitem{ArkaniHamed:1998rs}
N.~Arkani-Hamed, S.~Dimopoulos and G.~R.~Dvali,
{\emph{Phys.\ Lett.\ B} \textbf{429} (1998) 263-272},
[\href{https://arxiv.org/abs/hep-ph/9803315}{{\tt hep-ph/9803315}}].

\bibitem{Antoniadis:1998ig}
I.~Antoniadis, N.~Arkani-Hamed, S.~Dimopoulos and G.~R.~Dvali,
{\emph{Phys.\ Lett.\ B} \textbf{436} (1998) 257-263},
[\href{https://arxiv.org/abs/hep-ph/9804398}{{\tt hep-ph/9804398}}].


\bibitem{Guevara:2018wpp}
A.~Guevara, A.~Ochirov and J.~Vines,
{\emph{JHEP} \textbf{09} (2019) 056},
[\href{http://arxiv.org/abs/1812.06895}{{\tt 1812.06895}}].

\bibitem{Amati:1988tn}
D.~Amati, M.~Ciafaloni and G.~Veneziano,
{\emph{Phys.\ Lett.\ B} \textbf{216} (1989) 41-47}.


\bibitem{DiVecchia:2020ymx}
P.~Di Vecchia, C.~Heissenberg, R.~Russo and G.~Veneziano,
{\emph{Phys.\ Lett.\ B} \textbf{811} (2020), 135924},
[\href{http://arxiv.org/abs/2008.12743}{{\tt 2008.12743}}].

\bibitem{Bargheer:2013faa}
T.~Bargheer, J.~A.~Minahan and R.~Pereira,
{\emph{JHEP} \textbf{03} (2014) 096},
[\href{http://arxiv.org/abs/1311.7461}{{\tt 1311.7461}}].

\bibitem{Minahan:2014usa}
J.~A.~Minahan and R.~Pereira,
{\emph{JHEP} \textbf{04} (2015) 134},
[\href{http://arxiv.org/abs/1410.4746}{{\tt 1410.4746}}].

\bibitem{Antunes:2020pof}
A.~Antunes, M.~S.~Costa, T.~Hansen, A.~Salgarkar and S.~Sarkar,
{\emph{JHEP} \textbf{04} (2021) 088},
[\href{http://arxiv.org/abs/2012.01515}{{\tt 2012.01515}}].







\end{thebibliography}
\end{document}